\documentstyle[aps,epsfig]{revtex}
\begin{document}
\twocolumn[\hsize\textwidth\columnwidth\hsize\csname @twocolumnfalse\endcsname

\draft

\title{
Thermodynamics of doped Kondo insulator in one-dimension\\
- Finite-temperature DMRG study -
}

\author{
Naokazu {\sc Shibata} 
and Hirokazu {\sc Tsunetsugu}
}

\address{
Institute of Applied Physics, University of Tsukuba,
Tsukuba 305-8573, Japan
}

\maketitle
\begin{abstract}
The finite-temperature density-matrix renormalization-group 
method is applied to the one-dimensional Kondo lattice model
near half filling to study its thermodynamics. 
The spin and charge susceptibilities and entropy are 
calculated down to $T=0.03t$. We find two crossover temperatures near 
half filling.
The higher crossover temperature continuously connects to the spin
gap at half filling, and the susceptibilities are suppressed around 
this temperature.
At low temperatures, the susceptibilities increase again with
decreasing temperature when doping is finite. 
We confirm that they finally approach to the values obtained in the 
Tomonaga-Luttinger (TL) liquid ground state for several parameters.
The crossover temperature to the TL liquid is a new 
energy scale determined by gapless excitations of the TL liquid.
The transition from the metallic phase to the insulating phase 
is accompanied by the vanishing of the lower crossover temperature.
\end{abstract}

\vskip2pc]

\narrowtext


The Kondo insulator is a typical strongly correlated insulator 
and develops a spin gap at low temperatures\cite{Aeppli}.
The half-filled Kondo lattice (KL) model has been studied as its 
theoretical model, particularly in one dimension (1D) by
both numerical and analytical approaches\cite{THUS,YW,NU,Tsve,SNUI,FK}.
The ground state of this model is shown to have both
spin and charge gaps, 
$\Delta_{\mbox{\scriptsize s}}$, $\Delta_{\mbox{\scriptsize c}}$,
and strong correlation effects appear in their difference\cite{NU}.
The spin gap is always smaller than the charge gap, and
for small exchange coupling $J$ the spin gap is exponentially small,
$\Delta_{\mbox{\scriptsize s}}\sim \exp(-1/\alpha\rho J)$\cite{THUS,Tsve,SNUI,FK},
while the charge gap is linear in $J$\cite{NU,SNUI}.

At finite temperatures, the spin gap characterizes unique 
temperature dependence of the excitation spectrum. 
With increasing temperature, the structure of the charge gap 
in the dynamic charge structure factor and the quasiparticle 
density of states disappears at $T\sim\Delta_{\mbox{\scriptsize s}}$,
much lower than $\Delta_{\mbox{\scriptsize c}}$, although the latter 
is the energy scale of charge excitations at $T=0$\cite{Mutou,SU}.
This feature is also seen in the temperature dependence of the 
charge susceptibility, which drastically decreases below  
$T\sim\Delta_{\mbox{\scriptsize s}}$\cite{SATSU}.
As for the spin susceptibility, it decreases exponentially with the 
energy scale of $\Delta_{\mbox{\scriptsize s}}$, as expected\cite{Fye,SATSU}.

When a finite density of carriers are doped, the 1D KL model belongs 
to another universality class.
The ground state and the low energy excitations are described as a 
Tomonaga-Luttinger (TL) liquid\cite{UNT,fujimoto,MC,SNUI2,RMP,STU,YO}.
In contrast to the half filled case, the ground state has gapless 
excitations in both spin and charge channels.
Consequently the spin and charge susceptibilities are finite at $T=0$ and
determined by the velocities of the collective excitations,
 $\chi_{\mbox{\scriptsize s}}=1/(2\pi v_{\mbox{\scriptsize s}})$ 
and $\chi_{\mbox{\scriptsize c}}=2K_{\mbox{\scriptsize c}}/
(\pi v_{\mbox{\scriptsize c}})$ where $K_{\mbox{\scriptsize c}}$
is the Luttinger-liquid parameter\cite{Schul,Taka2,Usuki},
whereas  $\chi_{\mbox{\scriptsize s}}=0$, $\chi_{\mbox{\scriptsize c}}=0$
at half filling due to the gap.

In the present paper we study thermodynamics of the 1D
KL model near half filling, i.e. in the vicinity of 
 metal-insulator transition.
Thermodynamic quantities are calculated by the
finite-temperature density-matrix 
renormalization-group (finite-$T$ DMRG) 
method\cite{DMRG,Bursill,Wang,Shibata}, and
we find that the spin gap at half filling appears 
as a crossover temperature such that the susceptibilities
are suppressed around there.
We also find an new lower crossover temperature which is the
energy scale of the gapless excitations of the TL liquid 
ground state. This is sensitive to the hole doping,
and the transition from the metallic phase to the insulating phase
corresponds to the vanishing of the lower crossover temperature.

The Hamiltonian we use in the present study
is the 1D KL model 
described as
\begin{equation}
  {\cal H}= -t \sum_{i,s}(c^{\dagger}_{i s}c_{i+1 s}+\mbox{H.c.})
  +J\sum_{i,s,s'}{\bf S}_{i}\cdot{\textstyle\frac{1}{2}}
\mbox{\boldmath $\sigma$}_{ss'} c^{\dagger}_{is}c_{is'},
\label{KLM}
\end{equation}
where $\mbox{\boldmath $\sigma$}_{ss'}$ are the Pauli matrices
and ${\bf S}_i=\sum_{s,s'}\frac{1}{2}\mbox{\boldmath $\sigma$}_{ss'}
f^{\dagger}_{is}f_{is'}$ is the localized spin at site $i$.
The model has hoppings $-t$ ($t>0$) for only nearest neighbor pairs.
The density of conduction electron $n_{\mbox{\scriptsize c}}$ 
is unity at half filling,
and hole doping ($n_{\mbox{\scriptsize c}}=1-\delta$) is 
physically equivalent 
to electron doping ($n_{\mbox{\scriptsize c}}=1+\delta$) 
due to the particle-hole symmetry.

In order to study thermodynamics 
we employ the finite-$T$ DMRG method\cite{DMRG,Bursill,Wang,Shibata}.
By iteratively increasing the Trotter number
of the quantum transfer matrix, we can obtain the eigenvector
for the largest eigenvalue with desired accuracy.
Thermodynamic quantities are directly calculated
from this eigenvector,
and the extrapolation in the system size is not needed\cite{Betsu}.
This method was first applied to the quantum spin systems
and shown to be reliable 
down to the low temperature $T=0.01J$\cite{Wang,Shibata}. 
This method is free from statistical
errors and the negative sign problem, which are
advantages compared with the quantum Monte Carlo method.

To obtain thermodynamic quantities with fixed hole density 
$\delta=1-n_{\mbox{\scriptsize c}}$,
we need chemical potential $\mu$ at each temperature.
This requires many DMRG calculations at different fixed
chemical potentials: for each $J$ we use 36 sets of
$\mu$ with a typical interval $\Delta\mu = 0.025t$.
The number of states used in the present study
is typically 54 and corresponding truncation error is $10^{-3}$
at the lowest temperature $T=0.03t$ with the Trotter number 60.

\begin{figure}[t]
\epsfxsize=70mm \epsffile{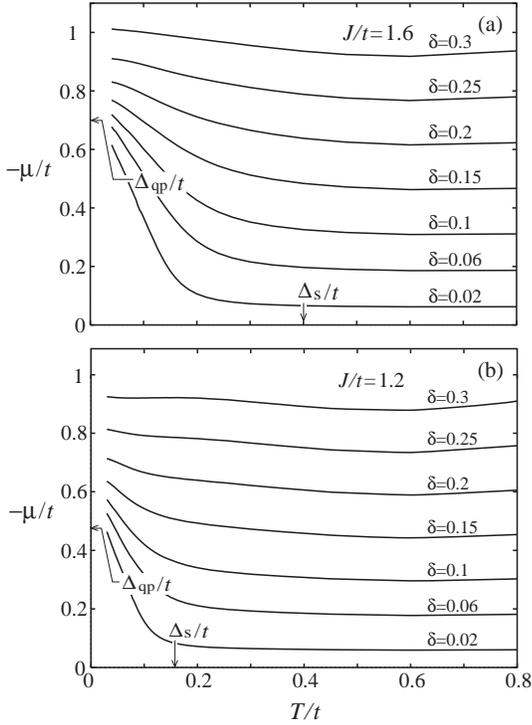}
\caption{
Chemical potential of the one-dimensional Kondo lattice model near
half-filling.
}
\end{figure}

The $T$-dependence of the chemical potential for several
$\delta$'s is shown in Fig.~1 for $J=1.6t$ and $1.2t$.
Note that the chemical potential at half filling is always
zero due to the particle-hole symmetry.
At high temperatures, $\mu$ is similar to the value for the 
free conduction electrons and indicates metallic behavior.
At low temperatures, however, a significant increase in $|-\mu|$ 
appears for small $\delta$.
In the limit of $T=0$, $\mu$ approaches $T=0$ value and this 
clearly shows the presence of the quasiparticle gap 
$\Delta_{\mbox{\scriptsize qp}}$ as shown in Fig.~1.
Previous $T=0$ DMRG calculations show that
$\Delta_{\mbox{\scriptsize qp}}=0.7t$ for $J=1.6t$ and 
$0.47t$ for $J=1.2t$, which is consistent with 
the present calculation.
In the following, we calculate thermodynamic quantities for fixed $\mu$'s
and convert them for fixed $\delta$'s using these data of $\mu(\delta,T)$.

Temperature dependence of spin susceptibility $\chi_{\mbox{\scriptsize s}}(T)$
is plotted in Fig.~2 for $J/t=1.6$ and $1.2$ at $0\le\delta\le 0.2$.
At high temperatures $\chi_{\mbox{\scriptsize s}}$ is 
asymptotically determined by the sum of the 
Curie term of the localized spins, $1/(4T)$, and the Pauli susceptibility,
$\chi_{\mbox{\scriptsize Pauli}}$, of the free conduction electrons. 
This is actually seen in the inset of Fig.~2(a).
For $\delta\leq0.2$, the change in the density 
of states of the free conduction band due to hole doping is within $5\%$
leading to little change in $\chi_{\mbox{\scriptsize Pauli}}$,
and it is reasonable that the $\delta$-dependence of 
$\chi_{\mbox{\scriptsize s}}$ is 
small at high temperatures.

With decreasing temperature the spin susceptibility increases 
owing to the Curie term of the localized spins, 
but around $T\sim\Delta_{\mbox{\scriptsize s}}$, the spin susceptibility 
starts to be suppressed as in the $\delta=0$ case.
Previous $T=0$ DMRG calculations show that
$\Delta_{\mbox{\scriptsize s}}=0.4t$ for $J=1.6t$,
and $\Delta_{\mbox{\scriptsize s}}=0.16t$ for $J=1.2t$\cite{SNUI}.
These behaviors suggest 
\begin{figure}[t]
\epsfxsize=70mm \epsffile{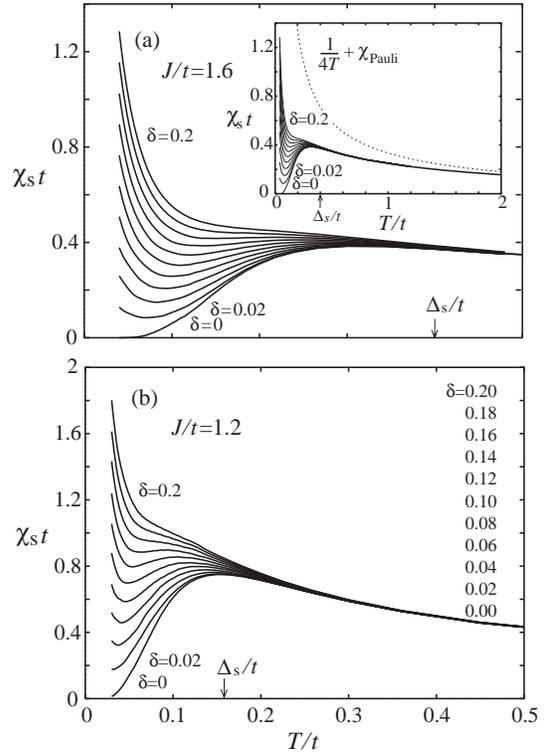}
\caption{
Spin susceptibility of the one-dimensional Kondo lattice model near
half-filling. We use the units of $(g\mu_B)^2=1$.
}
\end{figure}
\noindent
that the spin gap at $\delta=0$ persists as the 
crossover temperature characterizing 
the suppression of $\chi_{\mbox{\scriptsize s}}$ even away 
from half filling.

With further decreasing temperature, $\chi_{\mbox{\scriptsize s}}$ sharply 
increases again when doping is finite, 
whereas it exponentially decreases at half filling with the energy scale
of $\Delta_{\mbox{\scriptsize s}}$.
The increase in $\chi_{\mbox{\scriptsize s}}$ 
seems to be proportional to $\delta$ at low temperatures.
In order to see this $\delta$-dependence in more detail, 
we plot the difference in $\chi_{\mbox{\scriptsize s}}$ 
between $\delta>0$ and $\delta=0$ divided by $\delta$.
As shown in Fig.~3 for $J=1.6t$,
the universal behavior is observed at low temperatures indicating 
$\chi_{\mbox{\scriptsize s}}(T)\sim
\chi_{\mbox{\scriptsize s}}(T,\delta=0)+\delta/(4T)$. This means that 
the doped holes induce almost free spins of $S=\frac{1}{2}$ 
with density $\delta$.

In the limit of $T\rightarrow 0$, the thermodynamic properties 
are determined by the gapless collective excitations of the TL
 liquid, and the spin susceptibility is given by the spin velocity 
$v_{\mbox{\scriptsize s}}$ as $\chi_{\mbox{\scriptsize s}}=
1/(2\pi v_{\mbox{\scriptsize s}})$\cite{Schul,Taka2}.
Thus there must be a crossover temperature where 
$\chi_{\mbox{\scriptsize s}}(T)$ deviates from 
$\chi_{\mbox{\scriptsize s}}(T)\sim
\chi_{\mbox{\scriptsize s}}(T,\delta=0)+\delta/(4T)$
towards $1/(2\pi v_{\mbox{\scriptsize s}})$.
This crossover temperature is expected to be proportional to 
$v_{\mbox{\scriptsize s}}$, because it is the energy scale of 
the spin excitations. One can estimate this crossover temperature from
$\chi_{\mbox{\scriptsize s}}$ at $T=0$ through the 
relation $v_{\mbox{\scriptsize s}}=1/(2\pi \chi_{\mbox{\scriptsize s}})$.

The $\chi_{\mbox{\scriptsize s}}$($T$$=$$0$) are plotted in Fig.~4 for 
$J=2.0t$. They are calculated
by the zero-temperature DMRG method\cite{DMRG} with open boundary 
conditions via the size dependence of the lowest spin excitation 
energy, $\Delta E(L)=\pi v_{\mbox{\scriptsize s}}/L$.
The $T=0$ susceptibility is very sensitive to the 
\begin{figure}[t]
\epsfxsize=70mm \epsffile{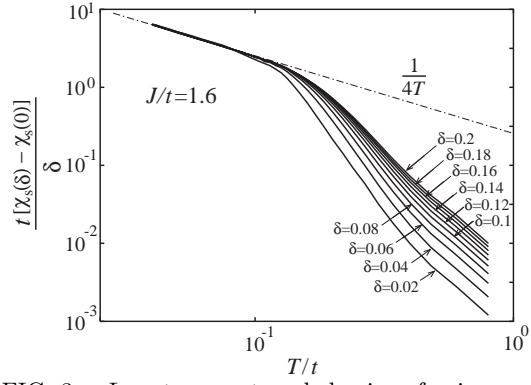}
\caption{
Low temperature behavior of spin susceptibility near half-filling.
}
\end{figure}
\begin{figure}[t]
\epsfxsize=67mm \epsffile{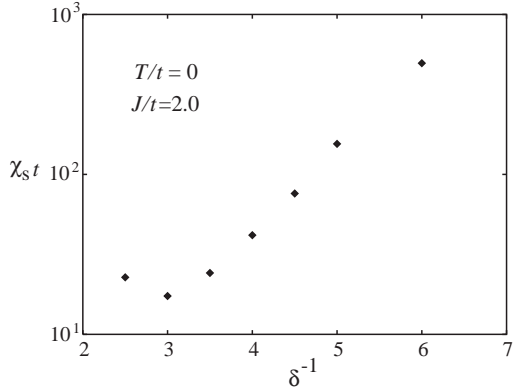}
\caption{
Doping dependence of spin susceptibility at $T=0$.}
\end{figure}
\noindent
hole doping $\delta$ and diverges in the limit of $\delta \rightarrow 0$.
It reflects the behavior that the spin velocity as a characteristic 
energy scale vanishes. Thus the crossover temperature 
to the TL liquid phase may correspondingly vanish as $\delta\rightarrow 0$.
The $\delta$-dependence of $\chi_{\mbox{\scriptsize s}}$ 
seems to be exponential in the present calculation,
but we will discuss this point again later.

The crossover to the TL liquid is
more clearly seen in the charge susceptibility, $\chi_{\mbox{\scriptsize c}}$.
The charge susceptibility is 
defined as the change in the conduction electron density
due to a small shift of chemical potential,
$\chi_{\mbox{\scriptsize c}}=\partial n_{\mbox{\scriptsize c}} 
/\partial \mu=-\partial \delta/\partial\mu$. 
The results for $J=1.6t$ and $1.2t$ are shown in Fig.~5.

At high temperatures, $\chi_{\mbox{\scriptsize c}}$ is almost the same as 
that of free conduction electrons and proportional to $1/T$.
With decreasing temperature, $\chi_{\mbox{\scriptsize c}}$ for small doping
is suppressed around $T\sim\Delta_{\mbox{\scriptsize s}}$, 
as for $\chi_{\mbox{\scriptsize s}}$.
The suppression of $\chi_{\mbox{\scriptsize c}}$ at $\delta=0$
is due to the development of the charge gap.
Below $T\sim\Delta_{\mbox{\scriptsize s}}$, the correlation length 
of the localized spins becomes longer than the charge correlation length,
and this induces an internal staggered magnetic field
for the conduction electrons through the Kondo coupling.
When the charge excitations are concerned, 
their time scale is much shorter than that for the spin excitations
($\propto\Delta_{\mbox{\scriptsize s}}^{-1}$), and the  
staggered field may be considered as almost 
static. This staggered field induces a unit-cell doubling for the 
\begin{figure}[t]
\epsfxsize=70mm \epsffile{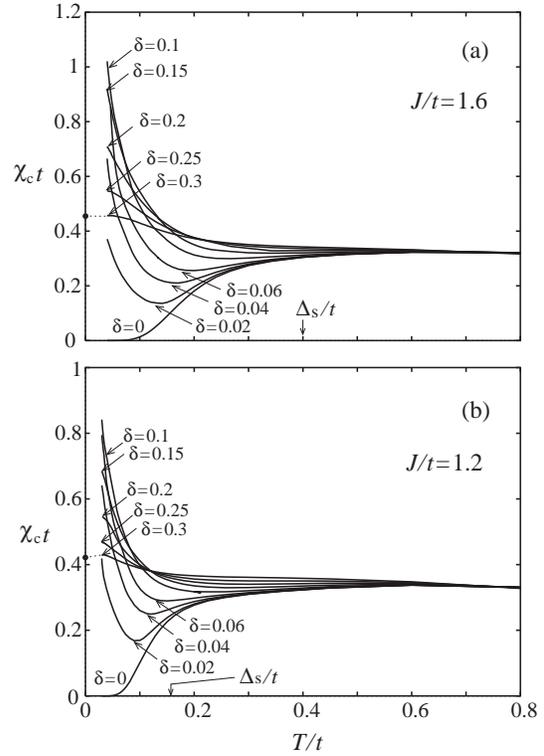}
\caption{
Charge susceptibility of the one-dimensional Kondo lattice model near
half-filling. The dot on the vertical axis is the $T=0$ value
calculated by the zero-temperature DMRG method.
}
\end{figure}
\noindent
conduction electrons and a finite gap appears 
at the Fermi energy of the otherwise free conduction
electrons. Since this behavior is induced by the development of 
long-range spin correlations, it becomes noticeable only below 
$T\sim\Delta_{\mbox{\scriptsize s}}$ rather than 
$\Delta_{\mbox{\scriptsize c}}$.
The present results show that contributions to 
$\chi_{\mbox{\scriptsize c}}$ 
from the small number of holes are small at temperature 
$T\sim \Delta_{\mbox{\scriptsize s}}$ and the suppression
is visible near half filling.

With further decreasing temperature, 
the temperature dependence of $\chi_{\mbox{\scriptsize c}}$ 
at $\delta>0$ differs from
the $\delta=0$ case as for $\chi_{\mbox{\scriptsize s}}(T)$. 
$\chi_{\mbox{\scriptsize c}}$ at half filling 
becomes exponentially small below $T\sim\Delta_{\mbox{\scriptsize s}}$
with the energy scale of the quasiparticle gap, 
$\chi_{\mbox{\scriptsize c}} \sim 
\exp(-\Delta_{\mbox{\scriptsize qp}}/T)$\cite{SATSU},
but for finite $\delta$, $\chi_{\mbox{\scriptsize c}}$ sharply 
increases.
The increase of $\chi_{\mbox{\scriptsize c}}$  
at low temperatures seems to be larger for smaller $\delta$.
This behavior near half filling is 
consistent with the doping dependence of 
$\chi_{\mbox{\scriptsize c}}$ at $T=0$, 
which is obtained by the zero-temperature DMRG method.
At $T=0$, $\chi_{\mbox{\scriptsize c}}$ is given by the difference in the
chemical potential $\chi_{\mbox{\scriptsize c}}=
2/L(\mu(L,N_{\mbox{\scriptsize c}}+1)-\mu(L,N_{\mbox{\scriptsize c}}-1))$,
where $2\mu(L,N_{\mbox{\scriptsize c}}+1)=
E_{\mbox{\scriptsize g}}(L,N_{\mbox{\scriptsize c}}+2)-
E_{\mbox{\scriptsize g}}(L,N_{\mbox{\scriptsize c}})$ and 
$E_{\mbox{\scriptsize g}}(L,N)$ 
is the ground state energy of $N$ conduction electrons in 
the system of length $L$. 
The results are shown in Fig.~6, which actually
shows that the $\chi_{\mbox{\scriptsize c}}$ increases with decreasing
$\delta$ and it seems to diverge in the limit of 
$\delta\rightarrow 0$. 
The divergent behavior of 
$\chi_{\mbox{\scriptsize c}}$ corresponds to the vanishing charge 
velocity as the characteristic energy of the TL liquid 
as $\delta\rightarrow 0$.
The $\delta$-dependence of $\chi_{\mbox{\scriptsize c}}$ 
is close to $1/\delta$ and different from the exponential
dependence observed 
\begin{figure}[t]
\epsfxsize=70mm \epsffile{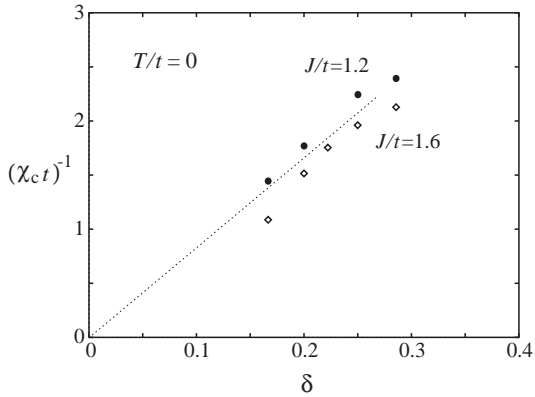}
\caption{
Doping dependence of charge susceptibility 
at $T=0$. The broken line is a guide for eyes.
}
\end{figure}
\begin{figure}[t]
\epsfxsize=70mm \epsffile{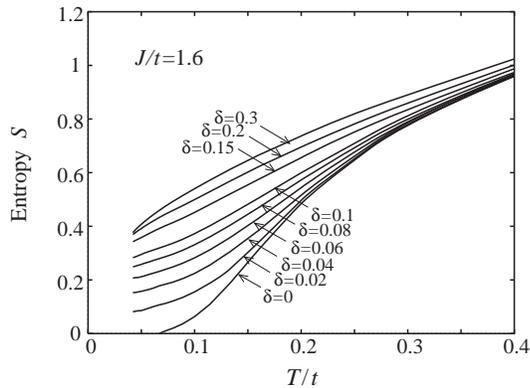}
\caption{
Entropy of the one-dimensional Kondo lattice model near
half filling.
}
\end{figure}
\noindent
in $\chi_{\mbox{\scriptsize s}}$ 
in the present DMRG calculation.
The $\chi_{\mbox{\scriptsize s}}$ seems to diverge stronger
than $\chi_{\mbox{\scriptsize c}}$ but this point is not clear
at the moment.

The new energy scale characterizing
gapless excitations of the TL liquid
may also be seen in the temperature dependence of entropy ${\cal S}(T)$.
The entropy is obtained as the derivative of the 
free energy with respect to temperature, and
the results are shown in Fig.~7 for $J=1.6t$.
At low temperatures the entropy at half filling becomes 
exponentially small reflecting finite spin and charge gaps.
When $\delta>0$, the entropy is enhanced at low 
temperatures.
This enhancement is due to the appearance of gapless 
excitations in the TL liquid.
Entropy of the TL liquid is proportional to $T$ at
low temperatures, ${\cal S}=\pi T(v_{\mbox{\scriptsize s}}^{-1}+
v_{\mbox{\scriptsize c}}^{-1})/3$\cite{Taka2,Usuki}.
However, for small $\delta$, the $T$-dependence is small even at 
temperatures $T\sim 0.05t$. 
The remaining entropy is consistent with the entropy of 
free spin-$\frac{1}{2}$ carriers with density $\delta$.
This implies that the characteristic energy scale for gapless 
excitations of the TL liquid is small near half filling and 
the $T$-linear dependence will be seen at further low temperatures
$T\ll 0.05t$.

In conclusion, we have applied finite-$T$ DMRG method to the 
one-dimensional Kondo lattice model away from half filling,
and found two crossover temperatures
which characterize thermodynamics near half filling. 
One is the spin gap at half filling and the other is the 
characteristic energy of collective excitations in the TL-liquid
ground state.
The energy scales of the TL liquid becomes smaller with approaching
half filling and
seems to vanish in the limit of $\delta\rightarrow 0$.

This work is financially supported by
Grant-in-Aid from the Ministry of Education, Science, Sports and
Culture of Japan.

\end{document}